\documentclass[aps,prl,showpacs,amsmath,amssymb,amsfonts,twocolumn,
superscriptaddress,floatfix,footinbib]{revtex4}

\usepackage{subfigure}
\usepackage{graphicx}
\usepackage[flushmargin]{footmisc}

\newcommand{\x}{\vec{x}}

\newcommand{\chix}{\vec{\chi}(\x)}
\newcommand{\phix}{\vec{\phi}(\x)}
\newcommand{\chixstar}{\vec{\chi}^*(\x)}

\begin{document}

\title{Interference of Degenerate Polariton Condensates in Microcavities}

 \author{T C H Liew}
 \author{A V Kavokin}
 \affiliation{School of Physics and Astronomy, University of
 Southampton, Highfield, Southampton SO17 1BJ, UK}
 \author{I A Shelykh}
 \affiliation{International Center for Condensed Matter Physics, Universidade de Brasilia, 70904-970, Brasilia-DF, Brazil}
 \affiliation{St. Petersburg State Polytechnical University, 195251, St. Petersburg,
 Russia}

\date{\today}

\begin{abstract}

\noindent We demonstrate theoretically that the interaction of two
degenerate condensates of exciton-polaritons in microcavities leads
to polarization dependent parametric oscillations. The resonant
polariton-polariton scattering is governed by the polarization of
the condensates and can be fully suppressed for certain
polarizations. A polarization controlled solid state optical gate is
proposed.
\end{abstract}

\maketitle

{\bf Introduction.} With the recent experimental confirmation of
Bose-Einstein condensation (BEC) in semiconductors
microcavities~\cite{Kasprzak}, the use of quantum interference to
construct new devices becomes a real possibility. The potentiality
of microcavities for the realization of micron-sized optical
parametric oscillators (OPOs) has been recently
revealed~\cite{Diederichs}. In these systems, photons in the cavity
are strongly coupled to excitons in the interior quantum wells,
forming new normal modes known as
(exciton-)polaritons~\cite{StrongCoupling}. Optical parametric
oscillations take place due to resonant polariton-polariton
scattering processes, which maintain coherence in the system.
Polaritons are optically injected in the initial (pump) state and
are then scattered into states with lower and higher frequencies,
referred to as signal and idler respectively. Since coherence is
preserved, the reverse scattering rebuilds the population of the
pump state allowing oscillations to develop in which polaritons
fluctuate between pump, signal and idler states.

\begin{figure}[h]
        \centering
        \includegraphics[width=8.116cm]{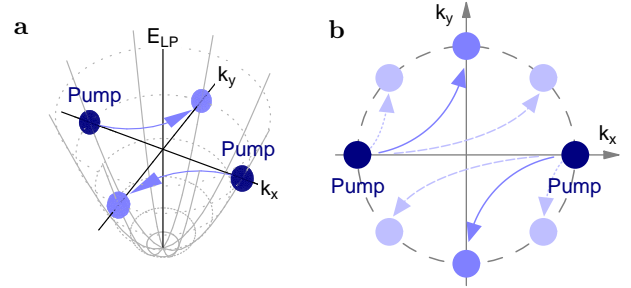}
    \caption{(color) Illustration of elastic scattering between two
    polaritons generated resonantly by pumps with equal and opposite
    wavevector. (a) the polariton dispersion;
    (b) the elastic circle in reciprocal
    space. Instead of being scattered uniformly around the elastic
    circle, as indicated by the lightest spots in (b), quantum interference causes preferential scattering to
    the wavevectors perpendicular to the pumps'.
    }
    \label{fig:Scheme}
    \end{figure}

Very recently, a degenerate OPO based on a microcavity has been
studied experimentally~\cite{Romanelli}. The cavity was pumped by
two continuous wave lasers having the same energy but opposite
in-plane wavevectors. In this configuration, the signal and idler
states are symmetric with respect to the pump states and have the
same energy (Fig. \ref{fig:Scheme}a), which is expected to allow for
much longer coherence times than in the conventional microcavity
OPO~\cite{Savidis,Saba}. In this Letter we present a quantum theory
of the degenerate OPO, which accounts for the polarization of light.
We show that the distribution of polaritons in reciprocal space and
their scattering are governed by the polarization of the two pumps,
so that a polarization controlled optical gate of a micron size can
be realized.

Polaritons are composite bosons and have two allowed spin
projections on the axis of the cavity ($\pm 1$). Their polarization
(linear, circular of elliptical) can be fully described by a 3D
vector called {\it pseudospin}~\cite{LarmorPrecession}.
Theoretically, the effect of polariton-polariton interactions is
accounted for using the zero-range interaction and mean-field
approximations that lead to the Gross-Pitaevskii equations. In
previous work these equations were used to study the scattering of
polaritons by a single impurity~\cite{Carusotto}, the spatial
structure formed by microcavity parametric oscillator
polaritons~\cite{Whittaker} and the dispersion of polariton
superfluids~\cite{Shelykh2006}. In this work we consider the
interference of polarized polariton liquids in the degenerate OPO.
From classical arguments one would expect polaritons to scatter
equally around an elastic circle in reciprocal space (Fig.
\ref{fig:Scheme}b). Indeed, all the scattering trajectories shown in
Fig. \ref{fig:Scheme}b are equally allowed by energy-momentum
conservation laws and the scattering matrix element can be assumed
wavevector independent with a good
accuracy~\cite{LarmorPrecession,Carusotto,Whittaker,Shelykh2006}.
However, experiments~\cite{Romanelli} and the theory presented here
reveal that a non-uniform intensity can appear. This symmetry
breaking of scattered polaritons takes place due to quantum
interference. Specifically, if the pumps are co-linearly polarized
then polaritons preferentially scatter to the wavevectors
perpendicular to the pumps' (Fig.\ref{fig:Scheme}a). Contrary, if
the pumps are cross-polarized then the scattering to perpendicular
directions is greatly suppressed.

{\bf Theoretical Model.} A coherent ensemble of polaritons is
described by two coupled wavefunctions, $\chix$ and $\phix$, which
represent excitons and photons in a microcavity respectively. Each
wavefunction is a vector with two components representing two
orthogonal linear polarizations (say $x$ and $y$). Following a
standard mean field treatment the evolution of the wavefunctions is
described by the Gross-Pitaevskii equations~\cite{Shelykh2006}:
    \begin{align}
    i\hbar\frac{\partial\chix}{\partial t}&=-\frac{\hbar^2\nabla^2}{2m_\chi}\chix+\Omega\phix+V_0\left(\chixstar\cdot\chix\right)\chix\notag\\
    &\hspace{10mm}-V_1\chixstar\left(\chix\cdot\chix\right)-\frac{i\hbar}{2\tau_\chi}\chix\label{eq:GPchi}
    \end{align}
    \begin{align}
    i\hbar\frac{\partial\phix}{\partial
    t}&=-\frac{\hbar^2\nabla^2}{2m_\phi}\phix+\Omega\chix+\vec{f}(\x,t)\notag\\
    &\hspace{10mm}+\vec{f_b}(\x,t)-\frac{i\hbar}{2\tau_\phi}\phix\label{eq:GPphi}
    \end{align}
\noindent $m_\chi$ and $m_\phi$ are effective masses assigned to the
parabolic dispersion of excitons and cavity photons with respect to
the in-plane wavevector. $\Omega$ is the exciton-photon coupling
energy, which is related to the quality factor and exciton decay
rate~\cite{CavityPolaritons,VCouplingConstant}. $V_0$ and $V_1$ are
constants determining the strength of the non-linear
interactions\cite{Shelykh2006}. $\tau_\chi$ and $\tau_\phi$ are the
lifetimes of excitons and photons, which account for the inelastic
scattering and radiative decay of polaritons. $f(\x,t)$ represents
an optical pumping, which for a single, continuous wave, Gaussian
pump is given by the Fourier integral:
    \begin{align}
    \vec{f}(\vec{x},t)&=\left(
    \begin{array}{c}A_x\\A_y\end{array}\right)\int\int e^{-iE_p
    t/\hbar}e^{-L^2(\vec{k}-\vec{k_p})^2/4}\notag\\&\hspace{5mm}\times\frac{i\Gamma}{\left(E_{LP}(\vec{k})-E_p-i\Gamma)\right)}d\vec{k}
    \label{eq:Pumpphi}
    \end{align}
\noindent $A_x$ and $A_y$ define the amplitudes of the two linearly
polarized components of the pump. $E_p$ is the pump energy, $k_p$ is
the pump in-plane wavevector and $L$ is the width of the laser spot
in real space. The pump energy should be chosen higher ($0.35$meV in
our calculation) than the energy of the bare lower polariton branch
to compensate for the energy renormalization (blue shift) caused by
forward scattering processes~\cite{WhittakerBlueShift}. The fraction
$\Gamma/ \left(E_{LP}(\vec{k})-E_p-i\Gamma)\right)$ is introduced to
account for non-uniform absorption of the Gaussian optical pump. The
fraction is derived from a Lorentz oscillator model where $\Gamma$
is the homogeneous oscillator (HWHM) linewidth~\cite{HaugKoch}.
$E_{LP}(\vec{k})$ is the bare dispersion of the lower polariton
branch. If there is more than one pump then $\vec{f}$ can be
constructed by a superposition of individual pumps.
$\vec{f_b}(\x,t)$ is a Langevin noise term much smaller in magnitude
($\times10^{-4}$) than the pump. We take this term as a randomly
changing white noise with no correlation between each point in space
and time. The noise term is necessary to seed the polariton
scattering in the case of exactly co-polarized pumps.

Eqs. (\ref{eq:GPchi}, \ref{eq:GPphi}) completely determine the
dynamics of interacting polaritons once initial wavefunctions and
parameters are defined. Note that in our model we have ignored the
longitudinal-transverse splitting of polaritons as it is not
essential in the experiment~\cite{Romanelli}. We also neglected the
disorder induced elastic scattering of polaritons, which is a linear
effect. It is dominated by the polariton-polariton scattering at
high pumping intensities.

\begin{figure}[t]
        \centering
        \includegraphics[width=8.116cm]{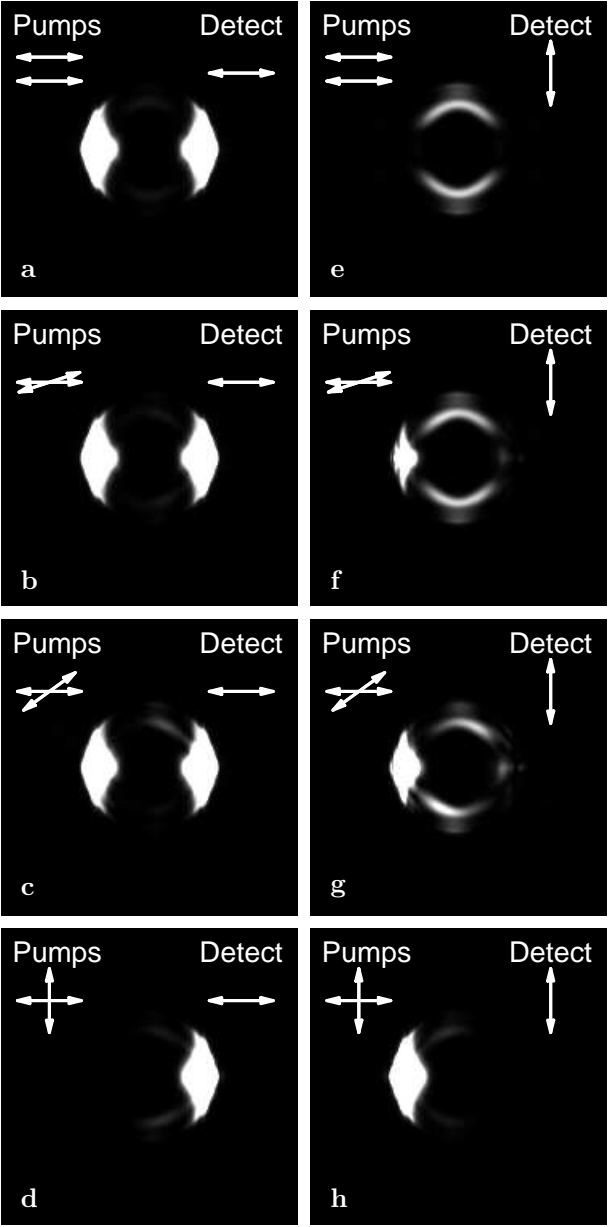}
    \caption{Time-averaged photon intensity in reciprocal space. (a-d) show the $x$ linearly polarized component; (e-h) show the $y$ polarized component. Different rows correspond to different pump polarizations; in (a, e) the pumps are co-linearly polarized, in (d, h) they are cross-polarized. Parameters: $m_\chi=0.22m_e$, $m_\phi=10^{-5}m_e$
    ($m_e$
    is the free electron mass), $\Omega=5.1$meV,
    $|\vec{k}_p|=785$mm$^{-1}$, $V_0=1.54\times10^{-5}$meV mm$^2$, $V_1=0.55V_0$, $\tau_\chi=10$ps, $\tau_\phi=1.6$ps, $|A_x, A_y|=190$meV mm$^{-1}$, $\Gamma=0.2$meV, $L=25\mu$m, $E_p=E_{LP}(\vec{k}_p)+0.35$meV.}
    \label{fig:kspaceInt}
    \end{figure}

\begin{figure}[t]
        \centering
        \includegraphics[width=8.116cm]{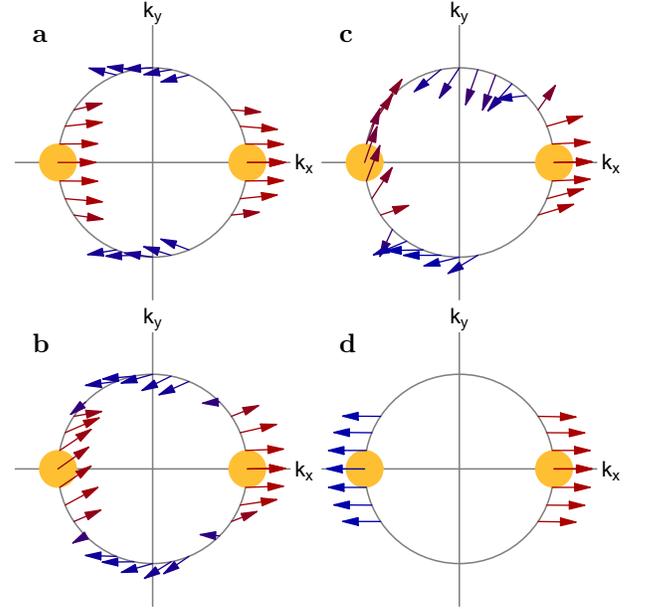}
    \caption{(color) The variation of the
    (time-averaged) pseudospin vector, $(S_x,S_y,S_z)$, around a circle in reciprocal space with
    radius equal to the pump wavevector. (a-d) correspond to the pump polarizations used before in Fig.\ref{fig:kspaceInt}.}
    \label{fig:kspacerho}
    \end{figure}

{\bf Microcavity Interference Device.} We consider the generation of
polaritons by two spatially overlapping linearly polarized pumps
with opposite in-plane wavevectors. We assume both pumps have the
same amplitude and phase. We solved Eqs. (\ref{eq:GPchi},
\ref{eq:GPphi}) numerically using a (fifth-order)
Adams-Bashforth-Moulton predictor-corrector method~\cite{Kincaid}.
The wavefunctions were initially set to zero at all points in space
and the evolution calculated for $2000$ps. Figure
\ref{fig:kspaceInt} shows the time integrated Fourier transform of
$\phix$, that is, the distribution of the photonic component in
reciprocal space. The photonic component of the polariton
wavefunction can be directly measured in optical experiments. When
the two pumps are co-linearly polarized there is strong scattering
to the wavevectors perpendicular to the pumps'. Note that this
result is in sharp contrast to what we expected classically. When
the two pumps are cross-linearly polarized the scattering to the
signal states (with wavevector perpendicular to the pumps') is
suppressed. The device thus functions as an optical gate where the
signal depends on the relative polarization of the input light.
Figure \ref{fig:kspacerho} illustrates the dependence of the
pseudospin vector around a circle in reciprocal space for different
mutual orientations of the two pump polarizations. The pseudospin
vector is the quantum analogue of the Stokes vector; its components
are given by:
\begin{equation}
S_x=\frac{|\phi_x|^2-|\phi_y|^2}{S_0}\hspace{5mm}S_y=\frac{\phi_x^*\phi_y+\phi_y^*\phi_x}{S_0}\notag
\end{equation}
\begin{equation}
S_z=i\frac{\phi_x^*\phi_y-\phi_y^*\phi_x}{S_0}\hspace{5mm}S_0=|\phi_x|^2+|\phi_y|^2\notag
\end{equation}
    \begin{figure}[h]
        \centering
        \includegraphics[width=8.116cm]{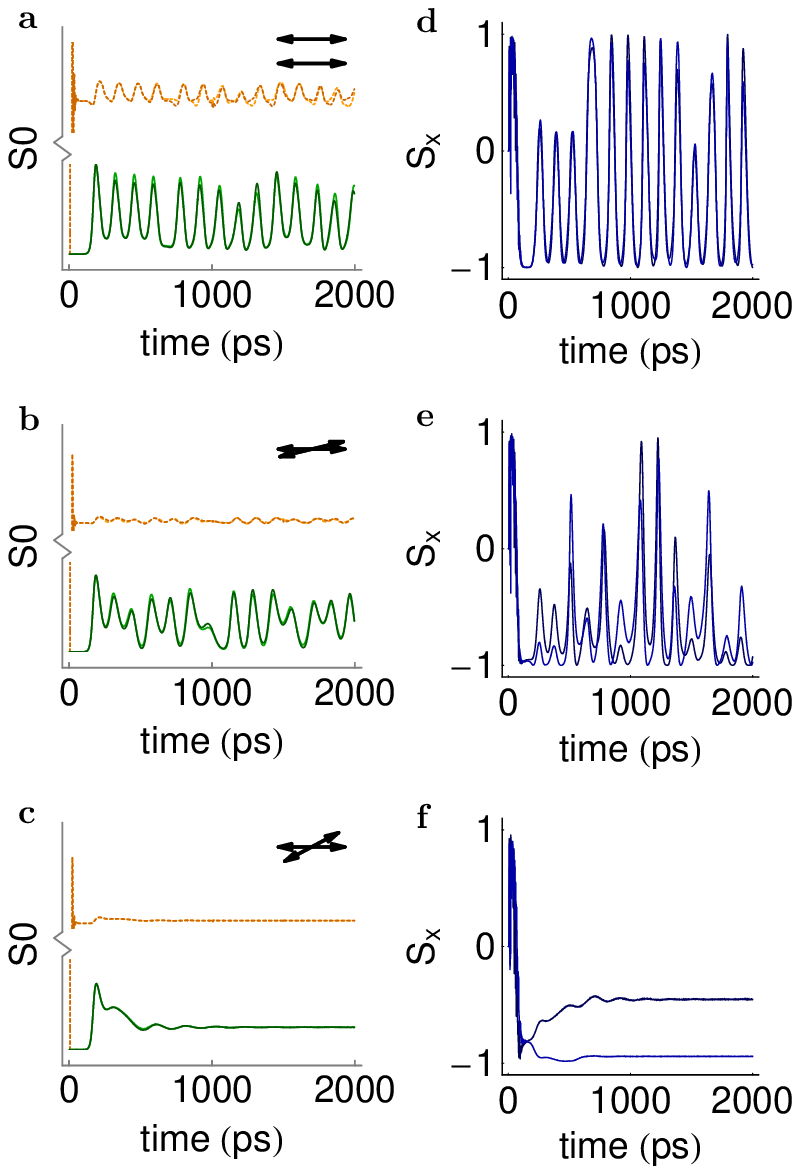}
    \caption{(color) Panels (a-c) show the time dependence of the total intensity,
    $S_0$, for the two signal states (lower solid curves) and the two pump
    states (upper dotted curves). Note that the scale on the $S_0$ axis is different in the lower and upper regions. Slight asymmetries appear between
    the two pump states and two signal states due to asymmetry in the Langevin noise. (a-c) use the same pump polarizations as in Fig.\ref{fig:kspaceInt} (a-c). Panels (d-f) show
    the corresponding evolution of the signal state horizontal
    polarization degree, $S_x$.}
    \label{fig:signaltime}
    \end{figure}
From the symmetry of the pump polariton pseudospins in the case of
cross-polarized pumps (Fig. \ref{fig:kspacerho}d) one sees that the
pseudospins of the signal states should not have any preferential
direction. From this we should expect the suppression of scattering
to the signal states. In the case of co-polarized pumps, we observe
stronger scattering to the cross-polarized states. This is a
consequence of the difference in sign between polariton-polariton
interaction constants in singlet and triplet configurations as
discussed in Ref. ~\cite{LarmorPrecession, Renucci2005}.

The signal state amplitudes also exhibit a non-trivial time
dependence (Fig. \ref{fig:signaltime}). In the case of co-polarized
pumps we can see an oscillatory behaviour with period
$T\approx135$ps. The oscillations become irregular and eventually
vanish if we increase the difference between the two pump
polarization directions. To better understand these oscillations we
use a 1D analytic model in which the exciton wavefunction on the
elastic circle in reciprocal space, $\chi(\theta)$, is parameterized
by the angle $\theta$. We neglect here the polarization degree of
freedom, the lifetime and the pump.
The Gross-Pitaevskii equation describing scattering dynamics is:
\begin{equation}
i\hbar\frac{\partial\chi(\theta))}{\partial
t}=\hbar\omega\chi(\theta)+V^\prime\chi^*(\theta-\pi)\int^\pi_{-\pi}\chi(\theta^\prime-\pi)\chi(\theta^\prime)d\theta^\prime\label{eq:GPAnalytic}
\end{equation}
\noindent where
$V^\prime=\left(V_0-V_1\right)|\vec{k}_p|^2/\left(16\pi^4\right)$
and $\hbar\omega$ is the (renormalized) exciton energy. Eq.
(\ref{eq:GPAnalytic}) accounts for the elastic scattering processes
where two excitons with wavevectors given by $\theta^\prime$ and
$\theta^\prime-\pi$ scatter to states with wavevectors given by
$\theta$ and $\theta-\pi$. Let us now introduce the variables:
\begin{align}
\rho(\theta)&=\rho(\theta-\pi)=\chi(\theta-\pi)\chi(\theta)\\
n(\theta)&=\chi^*(\theta)\chi(\theta)
\end{align}
\noindent The evolution of these variables can be derived from Eq.
(\ref{eq:GPAnalytic}). In the symmetric case,
$n(\theta)=n(\theta-\pi)$:
\begin{align}
i\hbar\frac{\partial\rho(\theta)}{\partial
t}&=2\hbar\omega\rho(\theta)+2V^\prime
n(\theta)\int^\pi_{-\pi}\rho(\theta^\prime)d\theta^\prime\label{eq:rho_ev}\\
i\hbar\frac{\partial n(\theta)}{\partial
t}&=V^\prime\left[\rho^*(\theta)\int^\pi_{-\pi}\rho(\theta^\prime)d\theta^\prime-\rho(\theta)\int^\pi_{-\pi}\rho^*(\theta^\prime)d\theta^\prime\right]\label{eq:n_ev}
\end{align}
\noindent Eq. (\ref{eq:n_ev}) shows that the total number of
excitons is conserved, that is, $\frac{\partial}{\partial
t}\int^\pi_{-\pi}n(\theta)d\theta=0$. To proceed we integrate Eq.
(\ref{eq:rho_ev}) with respect to $\theta$ and t, which gives:
\begin{equation}
\int^\pi_{-\pi}\rho(\theta)d\theta=A_0e^{-2i(\omega+V^\prime
N)t/\hbar}
\end{equation}
\noindent where $A_0=\int^\pi_{-\pi}\rho_0(\theta)d\theta$ and
$N=\int^\pi_{-\pi}n(\theta)d\theta$. Introducing the variables
$\tilde{\rho}(\theta)=\rho(\theta)e^{2i(\omega+V^\prime N)t/\hbar}$
and $t^\prime=2VNt$, we can now re-write Eqs. (\ref{eq:rho_ev},
\ref{eq:n_ev}):
\begin{align}
i\hbar\frac{\partial\tilde{\rho}(\theta)}{\partial
t^\prime}&=-\tilde{\rho}(\theta)+\frac{A_0 n(\theta)}{N}\\
i\hbar\frac{\partial n(\theta)}{\partial
t^\prime}&=\frac{A_0\tilde{\rho}^*(\theta)-A_0^*\tilde{\rho}(\theta)}{2N}
\end{align}
\noindent The solution of this set of coupled differential equations
is oscillatory and can be written in the form:
\begin{equation}
n(\theta)=B_0+B_1\sin\left(\frac{2\pi
t}{T}\right)+B_2\cos\left(\frac{2\pi t}{T}\right)
\end{equation}
where $B_0$, $B_1$ and $B_2$ are constants, determined by initial
conditions, and the period $T$ is:
\begin{equation}
T=\frac{\hbar \pi}{V^\prime\sqrt{N^2-|A_0|^2}}\label{eq:T}
\end{equation}
\noindent To compare this period to the numerical results we
calculate the constants $A_0$ and $N$ from the distribution of total
intensity, $S_0$, around the elastic circle. Eq. (\ref{eq:T}) then
estimates the period as $T=160$ps, which is close to the value
obtained numerically.

{\bf Conclusion.} We have demonstrated theoretically that the
quantum interference between polaritons created by two spatially
overlapping, degenerate, co-polarized laser beams, with opposite
in-plane wavevectors, results in enhanced scattering to
perpendicular directions. This symmetry breaking effect is in
contrast to what one expects from classical considerations. The
scattering is suppressed in the case of excitation with
cross-polarized lasers, giving the basis for a polarization
controlled solid state optical gate. Furthermore, the characteristic
parametric oscillations in intensity and polarization between pump
and signal states, that appear in the co-polarized case, can be
fully suppressed by increasing the difference between pump
polarizations.

The authors thank Yu. G. Rubo, M. M. Glazov, G. Malpuech, D.
Solnyshkov and N. A. Gippius for comments. I. A. Shelykh
acknowledges support from the Brazilian Ministry of Science and
Technology and IBEM, Brazil. T. C. H. Liew acknowledges support from
the E.P.S.R.C.


\end{document}